\documentclass[twocolumn]{revtex4}

\usepackage{ae}
\usepackage{psfrag}
\usepackage{graphicx}
\usepackage{amsmath,amssymb}

\newcommand{\rhoc}{\rho_{\scriptscriptstyle\rm c}}
\newcommand{\rhoa}{\rho_{\scriptscriptstyle a}}

\begin{document}
\title{Does mobility decrease cooperation?}

\author{\normalsize Mendeli H. Vainstein$^1$, Ana Tereza Costa Silva$^2$ and
  Jeferson J. Arenzon$^3$ \\ \normalsize
1. Instituto de F\'\i sica and International Center of Condensed \\ \normalsize
Matter Physics, Universidade de Bras\'\i lia \\ \normalsize
CP 04513, 70919-97, Brasília, DF, Brazil \\ \normalsize
2. Departamento de F\'\i sica, Universidade Estadual de Feira de 
             Santana \\ \normalsize Campus Universit\'ario, km 3, BR 116, \\ \normalsize
             44031-460 Feira de Santana BA, Brazil \\ \normalsize
3. Instituto de F\'\i sica, Universidade Federal do Rio Grande do Sul\\ \normalsize
         CP 15051, 91501-970 Porto Alegre RS, Brazil}
\date{}

\begin{abstract}
We explore the minimal conditions for sustainable cooperation on a spatially
distributed population of memoryless, unconditional strategies (cooperators and defectors)
in presence of unbiased, non contingent mobility in the context of the 
Prisoner's Dilemma game. We find that
cooperative behavior is not only possible but may even be enhanced by
such an ``always-move'' rule, when compared with the strongly viscous
(``never-move'') case. In addition, mobility also increases 
 the capability of cooperation to emerge and invade a population of
 defectors, what  may have a fundamental role in the
problem of the onset of cooperation.



\noindent
{\bf Corresponding author}: Jeferson J. Arenzon, +55 51 33166446
(phone), +55 51 33167286 (fax), arenzon@if.ufrgs.br

\noindent
{\bf Keywords:} game theory, cooperation, Prisoner's Dilemma

\end{abstract}

\maketitle


\section{Introduction}

The onset and sustainability of cooperation in social and non social populations
is still an open and challenging problem
\citep{Smith82,Axelrod84,DuRe98,HaSz05,SzFa06} that has been 
tackled with tools from different fields, ranging from psychological and social
sciences to statistical physics. Although involving a cost to the performer, 
cooperative behavior is ubiquitous in biological populations. Even more tantalizing is its
presence in groups of extremely simple
individuals~\citep{TuCh99,Crespi01,VuKo01,FrSc03,RaRa03,VeYu03,GrWeBu04,GrTr04,WoVaAr05,FiYuKaVe06,Mehdiabadi06},
where a mechanism other than direct or indirect reciprocity due to memory of previous encounters
or kinship relations should apply. Indeed, cooperative
behavior is found to occur when dispersal is very limited (high viscosity), what 
increases the probability of
future encounters  among close neighbors (the so called shadow of the future),
albeit decreasing the propagation rate of the strategies.
Axelrod~\citep{Axelrod84} was perhaps
the first to consider the effects of territoriality in the spread
of strategies in the Prisoner's Dilemma game (see definition below), 
either by colonization or imitation, but without explicit migration. 
Differently from the standard, random mixing population,
spatial localization allows a continuing interaction within the
local neighborhood.
The reasons for this are manyfold: individuals usually
occupy well-defined territorial regions, they do not move far from
their places of birth (population viscosity~\citep{Hamilton64}),
interactions occur in places where animals
usually meet such as water ponds, etc. That
preliminary study was later extended by Nowak and May~\citep{NoMa92,NoMa93}
who showed that geographical fixation enhances the probability of
further interaction in such a way that even simple 
nice rules like unconditional cooperation are able to survive. 
In these structured populations, cooperative strategies can build clusters
in which the benefits of mutual cooperation can outweight losses against
defectors, maintaining the population of cooperators stable.
These spatial games, where the interactions are localized and non random, have been
studied and extended in many ways (see, for example,
Refs.~\citep{NoBoMa94a,NoBoMa94b,LiNo94,Grim95,KiDo96,NaMaIw97,SzTo98,BrKiDo99,KiDoKn99,SzAnSzDr00,VaAr01,AbKu01,Hauert02,KiTrHoMiChCh02,Miekisz04,Aktipis04,HaDo04,FoVi05,SaPa05,DuMu05,EgZiCeMi05,SoMa06,SaPaLe06,HaAx06,Hauert06}).
Once the population is spatially structured, a
natural question concerns the effects of mobility that, along with other
important biological factors, is often neglected~\citep{Houston93}: is
it possible to evolve and sustain cooperation in a population of mobile agents,
where retaliation can be avoided by moving away from the former
partner? In particular, do we need explicit assortment, contingent
movements or any behaviorally complex strategy, or is it
possible to have a finite density of unconditional cooperators
with unbiased, random mobility? By increasing the effective range of interactions,
the introduction of mobility increases the random mixing and gets the system 
closer to the
mean field situation, in which every agent interacts randomly with
the whole population, and defection is known to prevail.
Thus, one might naively think that by dissipating 
the shadow of the future, mobility becomes a limiting factor for 
cooperation. 

Here we provide some insight on this issue by explicitly considering 
individual random diffusion in the framework of a locally, non 
randomly interacting spatial game, where simple, memoryless, 
strategy-pure agents coexist. 
This is important as it helps to settle the minimal
conditions under which cooperative behavior might emerge.
Although there is no simple answer to the
above question since motion can both destroy and enhance the
altruistic behavior, we show that there are broad conditions under which
even a blind pattern of mobility, without 
anticipating the future neighborhood (no assortment) and without 
considering the accumulated payoff, may have a
positive effect in the amount of cooperation. 
In other words, although mobility decreases the shadow of the future for 
nearest neighbors by diminishing the probability of a future encounter, 
it also increases it for more distant ones, that may now be visited.

Dugatkin and Wilson~\citep{DuWi91} and Enquist and Leimar~\citep{EnLe93}
showed that a randomly interacting population of fixed cooperators (playing Tit-for-Tat, 
TFT) could be invaded by mobile defectors that avoid retaliation
by moving in search of new cooperators to exploit. Mobility was introduced 
as a cost to wander between patches without spatial structure, not 
as an explicit diffusive process. By letting both mobility
and cooperative traits evolve together, Koella~\citep{Koella00}
(see also Ref.~\citep{BaRa98,HaTa05,LeFeDi05})
obtained low dispersive altruists and highly dispersive egoists which
enhanced the stability of local clusters. 
Again, there was no explicit diffusive behavior as
mobility was introduced by generating offspring within a given
dispersal range. Diffusion was considered by
Ferrière and Michod~\citep{FeMi95,FeMi96} by including a
diffusive term in the replicator equation~\citep{HoSi98}.
Two strategies, TFT and unconditional defection (D), were
allowed to move in a one dimensional system with local,
non random interactions, mobility again involving a cost.
This system may sustain cooperation when both strategies
have a minimum mobility, and retaliation by TFTs was found
to be an important ingredient.
More recently, Aktipis~\citep{Aktipis04} considered contingent movement of
cooperators: once a defection occurred in the previous movement,
they  walk away. This win-stay, lose-move strategy can invade a
population of defectors and resists further invasions.
Hamilton and Taborsky~\citep{HaTa05} and Le Galliard {\em et al.}~\citep{LeFeDi05}
(and Koella~\citep{Koella00} as well) considered the coevolution of mobility and cooperation
traits. However, both models are a kind of mean field approach as there
is no spatial structure and interactions are random. Models with alternating
viscosities, reflecting different stages of development that benefit both
from the clusterization of cooperators and dispersal, have also
been considered~\cite{WiPoDu92,Taylor92}, showing that local competition for
resources balances the benefits of kinship cooperation,  inhibiting
cooperation.
The present work differs from all these in several aspects: we consider
non random interactions on a two-dimensional structure, mobility traits
do not evolve and movements are Brownian, non contingent, and not  
under the control of the agents, both strategies considered are simple,  
unconditional and non retaliating, with no memory of previous steps. 
In other words, we are considering the simplest possible scenario
for cooperation.


We addressed in earlier work~\citep{VaAr01} the question of the
robustness of cooperation in spatial games in the presence of 
heterogeneous environments. By introducing quenched disorder 
in the lattice
(random dilution) each individual would sense a locally
varying social environment as the number of neighbors becomes
site dependent: optimal cooperation can be achieved
for weak disorder as the defects (or inaccessible regions)
act as pinning fields
for the strategy transition waves that cross the system,
keeping the clusters of cooperators more protected from
invasions. Thus, an irregular landscape may
enhance cooperation by introducing natural defenses
against invasions of defectors. Now we allow this disorder 
to be annealed: the vacant sites are no longer fixed and
may become occupied by a neighbor agent with a probability 
that depends on the populational viscosity. Only
random, unbiased diffusion is considered here, although extensions
to contingent
rules may be also devised. The detailed outcome of the
game will depend on the precise implementation of the
dynamics. For example,  the order in which combats,
offspring generation and diffusion occur leads to qualitative 
differences in the population.


\section{The Spatial Prisoner's Dilemma} 

The Prisoner's Dilemma game is
the archetypal model for reciprocal altruism.
In any round, each of the two players either cooperates (C) or
defects (D), without knowledge of the opponent's strategy.
The result depends on the mutual choice and is given by
the payoff matrix whose elements are:  a reward $R$ (punishment $P$)
if both cooperate (defect), $S$ (sucker's payoff) and $T$ (temptation)
if one cooperates and the other defects, respectively.
Moreover, these
quantities should satisfy the inequalities $T>R>P>S$ and
$2R>T+S$.
In a random mating, infinite population of asexual (haploid) elements,
where two pure strategies are present (cooperators $C$ and
defectors $D$), defecting will be the most rewarding strategy, independently
of the opponent's choice. 
Nonetheless, more complex rules (with memory of previous encounters) have been 
devised~\citep{Axelrod84} if the agents are
to meet again in the future. Here we will take a simplified version of the 
payoff matrix~\citep{NoMa92}:
$R=1$, $P=S=0$ and $T=b>1$, reducing the matrix to only one free parameter.
Initially, an equal number of cooperators and
defectors are randomly placed on a two dimensional square lattice of length
size $L$ and periodic boundary conditions, such that the total density
is $\rho$.
Each individual combats
with all its four closest neighbors (if any), accumulates the corresponding
payoff and then may either move or try to generate its
offspring. In the reproduction step,
each player compares its total payoff with the ones of its
neighbors and changes strategy, following the one with the
greatest payoff among them. This strategy changing updating rule preserves 
the total amount of individuals, thus keeping $\rho$ constant.
Results, averaged over 30-130 samples, are shown for $L=100$ and 
$b=1.4$, where the original
model ($\rho=1$) is known to sustain cooperation along with a finite fraction of strategy
changing, active sites.
As mentioned in the introduction, different values of $\rho$
can be used to mimic heterogeneous environments by allowing the
number of connections to vary from site to site due to dilution~\cite{VaAr01}.


There are several ways of implementing an unbiased random walk
along with the PD interactions. Here we consider two possibilities,
named combat-offspring-diffusion (COD) and combat-diffusion-offspring
(CDO). In the former, as the name says,  each step consists of  
combats followed by the generation of offspring done in parallel, and
then diffusion, while in the later, the diffusion and offspring
steps are reversed. During the diffusive step, each agent makes an
attempt to jump to a site chosen randomly within its four nearest
neighbours, what 
is accepted, provided the site is empty, with a probability $m$. Here
we only consider local steps with a reduced dispersal range (one lattice 
site), $m$ thus measuring the mobility of the agents ($m=0$
reduces to the case studied in \citep{VaAr01}).

\section{Results} 

\begin{figure}
\begin{center}
\psfrag{y}{\LARGE $\rhoc/\rho$}
\psfrag{t}{\LARGE $t$}
\includegraphics[width=8.5cm]{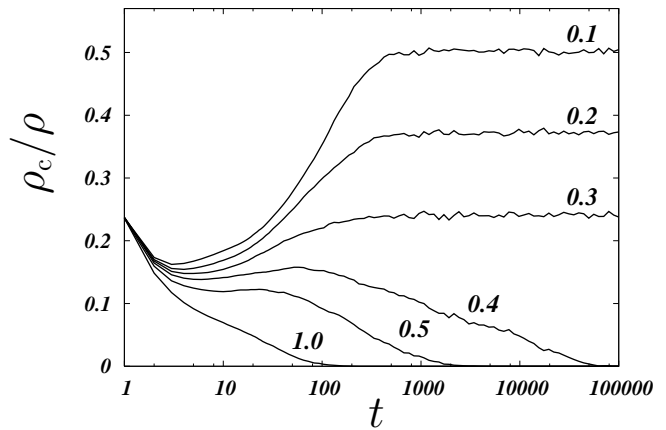}
\end{center}
\caption{Average fraction of cooperators $\rhoc/\rho$ as a function of 
time for several values of $m$ for the COD dynamics at $\rho=0.7$ in
a semi-log plot. In all cases there
is an initial decrease in the cooperators density since
cooperators are not yet coordinated and only form small groups,
being easily predated.
Depending on the mobility, at later times the existing
cooperator clusters may either disappear or grow,
leading to extinction or a stable, mixed population, respectively.
For values of $m\gtrsim 0.33$, mobility leads to extinction of
cooperation, even if sometimes very slowly. Indeed, close to the transition
point, the extinction time seems to diverge. On the other hand,
for low mobility, $m\lesssim 0.33$, after the initial decrease common
to all values of $m$, cooperation resumes and a plateau
is attained at intermediate values of $\rhoc$, with cooperators and
defectors coexisting. Notice
also that for $m=0$, $\rhoc/\rho\simeq 0.2$ (not shown)~\citep{VaAr01}:
the behavior, for $m=0$ and $m\to 0^+$, is quite different.}
\label{fig.COD_rhoctime}
\end{figure}

\begin{figure}
\begin{center}
\psfrag{y}{\LARGE $\rhoc/\rho$}
\psfrag{t}{\LARGE $t$}
\includegraphics[width=8.5cm]{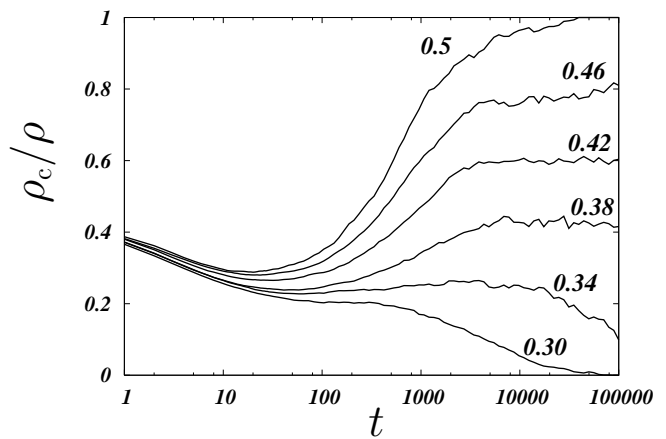}
\end{center}
\caption{Average fraction of cooperators $\rhoc/\rho$ as a function of 
time for $\rho=0.24$ and several values of $m$ for the CDO dynamics near the
transition from an all-D to an all-C phase. Analogously to fig.~\ref{fig.COD_rhoctime},
after the common initial decrease in the amount of cooperation,  at late times the existing
cooperator clusters may either disappear or increase their sizes, depending on the mobility $m$.
On the other hand, in this case cooperators may fully invade the
population and $\rhoc=\rho$.}
\label{fig.CDO_rhoctime}
\end{figure}

Figs.~\ref{fig.COD_rhoctime} and \ref{fig.CDO_rhoctime},
where the average temporal evolution of the cooperators density
$\rhoc$ is shown
for different values of the viscosity parameter $m$, exemplify the rich 
behavior presented by the model once mobility is introduced. Under
thinning or thickening, the ultimate fate of a population depends on the total 
density (and probably on the initial state), as is exemplified in 
these figures:  while in the COD dynamics of fig.~\ref{fig.COD_rhoctime}
the asymptotic density of cooperators decreases as $m$ increases, in
the CDO case of fig.~\ref{fig.CDO_rhoctime}, on the contrary,  
$\rhoc$ may increase with $m$ for some values of $\rho$. The
short time behavior is similar in both cases: the density of cooperators 
initially decreases since they are not yet coordinated (the initial state
is random) forming only small groups, what does not prevent the exploitation
by neighboring defectors.

\begin{figure}
\begin{center}
\psfrag{rhoc}{\LARGE $\rhoc/\rho$}
\psfrag{rhoa}{\large $\rhoa/\rho$}
\psfrag{rtot}{\LARGE $\rho$}
\includegraphics[width=8.5cm]{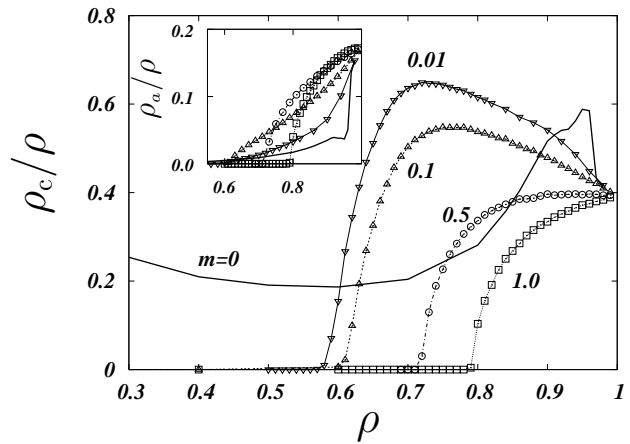}
\end{center}
\caption{Average fraction of cooperating individuals $\rhoc/\rho$ and active sites
$\rhoa/\rho$ (inset) for the COD dynamics and different values of the mobility, 
from $m=0$ (solid line) to 1. Different from the CDO case
(see Fig.~\ref{fig.CDO_rhoc}), here there is not a defector-free phase 
($\rhoc/\rho$ is always lower than 1) at intermediate densities, although
their presence is reduced. For small mobilities (smaller than 0.5 in the
figure) there is an optimal density where the relative amount of cooperators is
maximized. This maximum increases as $m$ decreases. Notice that as soon as there are
cooperators, there are also active sites: there is no frozen mixed configurations
in this case. Thus, as was exemplified in fig.~\ref{fig.COD_rhoctime}, for a
fixed density, a very tiny mobility is usually the best scenario at intermediate
densities. For small and high values of $\rho$, the viscous, ``never-move'' case
performs better. In particular, the $m=1$ always has less cooperators than the
immobile case ($m=0$). In the inset, the corresponding fraction of active
sites are plotted: the mixed state where Cs and Ds coexist is also an active
phase. Differently from the $m=0$ case, mobility, even if in small amounts,
helps to unpin the strategy-flipping waves that roam the system.}
\label{fig.COD_rhoc}
\end{figure}

\begin{figure}
\begin{center}
\psfrag{rhoc}{\LARGE $\rhoc/\rho$}
\psfrag{p}{\LARGE $m$}
\includegraphics[width=8.5cm]{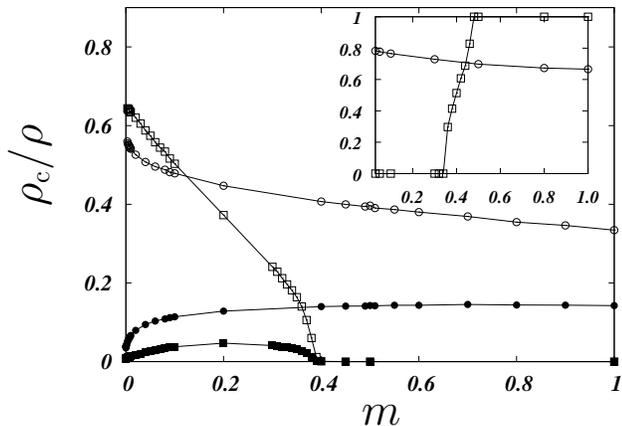}
\end{center}
\caption{Average fraction of cooperating individuals (empty symbols) and active 
sites (filled symbols) for the COD dynamics and densities $\rho=0.7$ (squares) and
0.9 (circles). For comparison, the values of $\rhoc/\rho$ for $m=0$ are:
0.2 ($\rho=0.7$) and 0.52 ($\rho=0.9$). For all values of $\rho$, the amount of
cooperation is a decreasing function of the mobility. Whether cooperative behavior
exist depends, however, on the value of $\rho$: while for $\rho=0.9$ cooperators
exist in the whole region, for $\rho=0.7$ a critical value of $m$ (around
0.33) exists above which defectors dominate. In analogy with some physical systems,
the dynamics close to this point slows down and long time simulations are
needed in order to extract the correct location and order of the transition.
Inset: the same as above but for the CDO dynamics. Notice that in
this case, the cooperation {\em increases} with $m$ for some values of $\rho$ (e.g.,
0.24, squares), and decreases for others (0.8, circles).}
\label{fig.COD_rho0.7}
\end{figure}

Fig.~\ref{fig.COD_rhoc}, for COD dynamics, shows both the density of 
cooperators and strategy-changing individuals (active sites, $\rhoa$), as a
function of the total lattice occupation $\rho$,  after the system
attained a stationary state where both quantities fluctuate
around their average values. Also shown, for comparison, are the results from 
\citep{VaAr01} for the extremely viscous case $m=0$. Cooperation
only appears above a minimum, $m$-dependent, density; below this
point,  defectors dominate ($\rhoc=0$). 
At low densities, any mobility
destroys cooperation ($\rhoc=0$): for $m=0$, isolated all-cooperating 
clusters are able to survive, but as soon as $m>0$ the existence
of free riders will invade these small clusters.
Although cooperation is possible for large mobilities (e.g., $m=1$),
cooperators perform better when nobody moves,
$\rhoc(m=0)>\rhoc(m=1)$, for all $\rho$.
Interestingly, for intermediate values of the mobility
(e.g., $m=0.1$) cooperation is enhanced when
compared with the viscous case: for a broad range of
densities, $\rhoc(m=0.1)>\rhoc(m=0)$. Thus, two immediate
conclusions are: first, cooperation is possible in the presence
of mobility when the available space is somewhat reduced and,
second, intermediate mobilities enhance cooperation!
Indeed, for intermediate mobilities, there is a maximum
in the fraction of cooperators (e.g., for $m=0.1$, the
maximum occurs at $\rho\simeq 0.75$), differently from
the $m=1$ case where this maximum occurs at $\rho=1$, 
where no movement is allowed. The transition from the
region with $\rhoc=0$ to the cooperative one seems to be continuous
and the finite fraction of active sites indicates that when $\rhoc\neq 0$,
both strategies, C and D, coexist.
Remarkably, the fall of cooperation after the maximum seems not to be 
associated with any particular behavior of active sites, whose fraction keeps growing
with the total density. Thus, although no particular sign is observed
in $\rhoa$ around the maximum of $\rhoc$, the decrease of cooperation after the
maximum is related to a smaller number of empty sites that act as pinning points that 
slow the dynamics or even prevent that some regions of cooperators be
predated, as was observed in \cite{VaAr01}.  Fig.~\ref{fig.COD_rho0.7} shows
the fraction of cooperators as a function of $m$ for two different
densities, 0.7 and 0.9. For both densities, mobility decreases cooperation,
and in the former, even destroys it completely above a threshold
 (close to it, the relaxation becomes too slow
and longer runs should be performed in order to decide whether the transition
is continuous or not). Both cases also differ on the role played by
the active sites, much more prominent for $\rho=0.9$ because the smaller
the number of empty sites (pinning points), the larger the number of active sites.

\begin{figure}
\begin{center}
\psfrag{rhoc}{\LARGE $\rhoc/\rho$}
\psfrag{rtot}{\LARGE $\rho$}
\includegraphics[width=8.5cm]{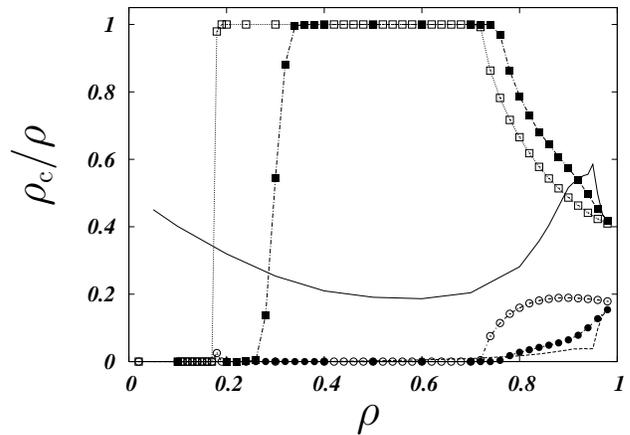}
\end{center}
\caption{Average fraction of cooperating individuals, $\rhoc/\rho$ (squares), and active 
sites (circles), $\rhoa/\rho$,
for the CDO dynamics and several values of the mobility: $m=0$ (solid
line), 1 (hollow symbols)
and 0.1 (filled symbols). Notice the three regimes: defector dominated
($\rhoc=0$, at low $\rho$), cooperator dominated ($\rhoc=1$, at
intermediate $\rho$) and a mixed one ($0<\rhoc<1$, at greater
values of $\rho$). They are
separated by two transitions, discontinuous and continuous, respectively.
Notice also that active sites, those that change strategy  at a
given time, have non neglectable densities only at large densities (roughly
above $\rho\simeq 0.7$), where there is a mixed phase with both strategies
coexisting ($0<\rhoc<\rho$). Again, in analogy with the COD
case, mobility enhances cooperation for intermediate values of
densities. However, higher mobilities increase the range of $\rho$
of the all-C phase.}
\label{fig.CDO_rhoc}
\end{figure}

\begin{figure}
\begin{center}
\psfrag{rhoc}{\hspace{-1.2cm}\LARGE Prob($\rhoc=1$)}
\psfrag{r}{\LARGE $\rho$}
\includegraphics[width=8.5cm]{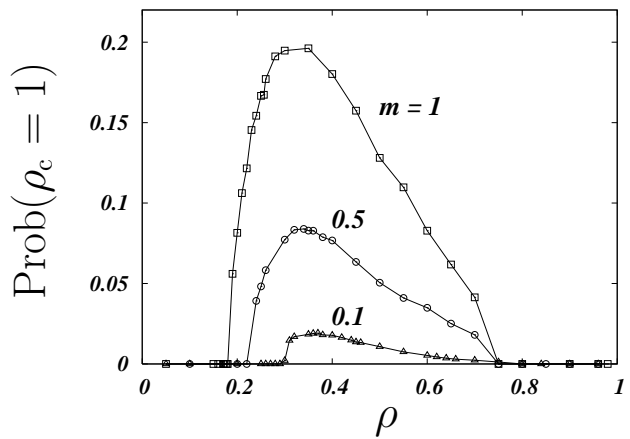}
\end{center}
\caption{Probability of invasion of a population of defectors by
cooperators, as a function of $\rho$, when an initial patch with
two cooperators is immersed in a whole population of defectors.
We consider several values of $m$ and CDO dynamics. 
Notice that the larger the mobility is, the greater is
the probability of invasion. This effect is absent in the corresponding 
viscous case ($m=0$), where this small patch is not enough for invasion to occur.}
\label{fig.CDO_2c}
\end{figure}

Fig.~\ref{fig.CDO_rhoc} presents the long time behavior of the
density of cooperators shown in fig.~\ref{fig.CDO_rhoctime} for the CDO dynamics,
as a function of the
total lattice occupation $\rho$, along with the fraction of active sites $\rhoa$.
Again, the overall picture remains the same:  mobility
destroys cooperation for low densities, while enhances it 
for higher densities. This effect is even stronger here than
in the COD case: besides occuring in a wider range of $\rho$ 
(compare figs.~\ref{fig.COD_rhoc} and \ref{fig.CDO_rhoc}), cooperators can 
invade completely the population ($\rhoc=\rho$, for some $\rho$, in 
fig.~\ref{fig.CDO_rhoc} and $\rhoc<\rho$, for all $\rho$, in fig.~\ref{fig.COD_rhoc}). 
Also, when compared with the viscous
$m=0$ case, this dynamics outperforms it, except very close to
$\rho=1$.  The origin of
the difference between CDO and COD dynamics is that, whenever $m\neq
0$, it is always good
for the cooperators to move away from its partner, whatever
its strategy, what favors the CDO dynamics. 
Differently from the previous case, here there are
two transitions: a sharp one from a D-dominated ($\rhoc=0$) to 
a C-dominated ($\rhoc=\rho$), followed by a continuous one
to an active phase (both strategies coexist and $\rhoa\neq 0$).
Moreover, for a given $\rho$, the dependence on mobility
is more complex than the previous case, as shown in the
inset of Fig.~\ref{fig.COD_rho0.7}: for large densities,
the behavior is analogous to the COD dynamics, while the
behavior for intermediate densities is unexpected, as
the system passes from an all-D to an all-C state as
$m$ increases. Although we do not deal in this paper
with the question of invadability of a population by
a different strategy, we present in Fig.~\ref{fig.CDO_2c},
an example of when an initial patch with only two cooperators
completely replaces the sea of defectors in which they are
immersed. Again, in this case the
mobility enhances the effect (unless the density is so high
that movements are prevented), and the larger $m$ is, the
greater is the probability of cooperators to invade the
population. In comparison, this
has a very small probability of happening when $m\simeq 0$.

\section{Discussion and Conclusions}

High population viscosity, or very limited dispersal (low mobility), is 
a possible mechanism for the emergence and 
maintenance of cooperation, even in a population of very simple, 
non-retaliating, strategy-pure, agents~\citep{NoMa92}. The
cluster organization prevents defectors from completely overtaking the
population because the payoff from the bulk cooperators outwin the
exploitation at the borders. A fundamental problem consists in obtaining
the minimal conditions under which cooperation is present in a 
population. In particular, in this paper, we tried to shed some
light in the role of mobility, a usually neglected factor. On one
hand, besides helping to spread clusters of cooperators, mobility may 
allow defectors to escape retaliation from
a former partner and helps to increase the random mixing of a
population by increasing the range of interaction.  Once
memoryless agents are considered (no recognition process is involved), the 
probability of future encounters (the so called shadow of the future) increases when the mobility is
small, and spatial correlations are strong.
Moreover, when increasing the mobility, the effective range of interactions 
increases proportionally to it and the probability of sharing the same opponent
decreases, thus dissipating the shadow of the future. In this case,
defection is expected. On the other hand, contingent mobility is expected to
enhance cooperation by avoiding continued exploitation or defector
rich regions. In between, when diffusion is unbiased  and the 
strategies, unconditional, it is not obvious what would be the
effect of mobility.

Here we presented results for
a simple spatial game where the patched environment allows
explicit, although random, movement of agents whose
strategies are pure, non retaliating (unconditional). The
diffusion is brownian,  not relying on any type of explicit,
genotypic or phenotypic assortment. Moreover, in our model
there is no correlation between mobility and altruism: all rules are
equally mobile.
Whether a given strategy is able or not to invade another population  
would strongly depend on how viscous a population is, the global
density and the chosen dynamics. However, some universal conclusions
can be stated. First of all, cooperation is possible under the
above conditions, thus enlarging the limits for cooperative 
behavior. Second, for a broad range of the parameters (density,
viscosity, etc), cooperation is enhanced in respect to the
viscous case. Third, a rule like always-move, regardless of
the opponent strategy, may increase the capability of
cooperators to invade and overtake a population of defectors.
In this sense, mobility may have a fundamental role in the
problem of the onset of cooperation.
Once mobility is incorporated within a population, it may evolve to
more contingent forms, perhaps under the control 
of the agents, and be strategy, payoff or partner dependent.
A possible realization of such diffusion scenario may
occur in organisms with extracellular metabolism, as is
the case of some  yeast cells ~\cite{GrTr04}.
Sugar is processed outside the cell by a secreted enzyme called invertase,
creating a common resource for all 
surrounding cells. This offers the opportunity for defection as
some cells may not have the cost of producing the enzyme but
yet benefit from that produced by others. To what extent the cooperative 
behavior observed in such simple organisms is a sole effect of the underlying 
spatial structure or whether there is an enhancement factor due to diffusion 
is an open and interesting question. Moreover, in systems that present
polarized motion (chemotaxis),  depending on the concentration
of cooperators a gradient of nutrients may be present and both cooperators 
and defectors can migrate towards (away) high cooperator (defector) density 
regions. Thus, a possible rule in this case can be: ``cooperators 
attract--defectors repell'', but many others can be devised.
Such non random rules may also develop cooperative swarming, 
relevant for evolving higher levels of biological organization, 
and is already found to occur in bacteria~\cite{VeYu03}.
Also interesting is the effect of these more
complex dynamical rules as well as the consequences of
random mobility in other regions of the strategy space, in particular,
when involving TIT-FOR-TAT players.

The mobility $m$ is an intrinsic parameter that
indicates the individual capability of performing walks (with unitary step). 
As so, data for the same density $\rho$ but different mobilities $m$
can be directly compared in figs.~\ref{fig.COD_rhoc} and
\ref{fig.CDO_rhoc}, and this information is summarized in
fig.~\ref{fig.COD_rho0.7}.
However, this parameter alone is not a measure of the 
effective dispersal since, as the density increases, 
movements are prevented by the lack of empty space. Unfortunately
there is not a general prescription for this effective dispersal
and an actual measure
would be necessary. This is an important point, that will be
addressed in a forthcoming publication along with the question of 
how the diffusivity of
individuals change, for a given $m$ and $\rho$, when 
changing the displacement rule from the random case considered
here to a more biased choice? Even for the simple unbiased case
of this work some preliminary results indicate that the effective
mobility is not a simple function of these parameters as the
dynamics may become very slow, due both to the presence of defects
or critical slowing down, analogous to those observed in glassy systems
and close to a continuous transition, respectively.

The PD is not the only possible framework in which social dilemmas
can be studied. For example, the snowdrift game~\cite{SzFa06,DoHa05}, where
$P<S$ (while $P>S$ in the PD), is also biologically relevant
and may lead to persistent cooperative behavior. The payoff matrix considers 
that cooperation involves a benefit $b$ to those involved and a cost $c$ 
 to the performer, while defection involves no costs or benefits. When both cooperate, 
they receive $R=b-c/2$, sharing the cost, while if they both defect they receive $P=0$.
When one cooperates and the other defects, the later receives $T=b$ while the
former is penalized with the total cost $S=b-c$.  Although $b>c>0$ and $P<S$, if we allow higher
costs, $b<c<2b$, we have $P>S$, recovering the PD ranking. Interestingly, when taking
the spatial structure into account in the snowdrift game, the amount of cooperation
may be reduced, depending on the cost to benefit ratio $c/(2b-c)$~\cite{HaDo04}.
It would be interesting to extend our results and 
study the effects of dilution and mobility in the snowdrift game, and different
updating rules,  stochastic or deterministic, synchronous or not, as well.
The parametrization considered in our work, proposed in the original
work of Nowak and May~\cite{NoMa92}, is at the borderline between these two
games as 
it considers $b=c$. There are, however, other possible one-parameter
matrixes, still keeping the ranking of the PD game, for example,
$T=1+r, R=1, P=0$ and $S=-r$, with $r=c/(b-c)$. 







{\bf Acknowledgments:} This work was partially supported by the Brazilian agencies CAPES,
CNPq and FAPERGS. MHV acknowledges the Department de Física Fonamental
of the Universitat de Barcelona where part of this work was
developed. ATCS acknowledges the hospitality of the IF-UFRGS during her visit
where part of this work was done.

\bibliographystyle{jtbnew}


\end{document}